\documentclass[galaxies,review,submit,pdftex,oneauthor]{Definitions/mdpi} 

\firstpage{1} 
\pubvolume{1}
\issuenum{1}
\articlenumber{0}
\pubyear{2024}
\copyrightyear{2024}
\datereceived{ } 
\daterevised{ } 
\dateaccepted{ } 
\datepublished{ } 
\hreflink{https://doi.org/} 

\Title{A review of long lasting activities of the central engine of gamma-ray bursts}

\TitleCitation{Long lasting activities of GRBs central engines}

\Author{Bruce Gendre $^{1, 2}$\orcidA{}}

\AuthorNames{Bruce Gendre}

\AuthorCitation{Gendre, B.}

\address{%
$^{1}$ \quad The University of the Virgin Islands, College of Science and Maths, 2 John Brewer's Bay, 00802, St Thomas, VI, USA\\
$^{2}$ \quad The University of Western Australia-OzGrav, 35 Stirling Highway, 6009, Crawley, WA, Australia}

\corres{Correspondence: bruce.gendre@uvi.edu}

\abstract{Gamma-ray bursts are known to display various features on top of their canonical behavior. In this short review, we will describe and discuss two of them: the ultra-long gamma-ray burst, which are defined by an extreme duration of their prompt phase, and the plateau phase, which is defined by a steady phase of large duration at the start of the afterglow. We will review the main properties of those two phenomena, and will discuss their possible origin, in light of the standard fireball model of gamma-ray bursts. A final section will discuss the future missions which could bring new evidences to the study of those objects.}


\begin{document}


\section{Introduction}

Etymologically, Gamma-Ray Bursts (GRBs) are observationally defined events: they are bursts of $\gamma$-ray photons. The phenomenon was discovered in the late 1960's by the Vela satellites \cite{kle73}, and since then has been studied by various dedicated instruments (e. g. KONUS, PHEBUS, BATSE, BeppoSAX, HETE-2, Swift,\cite{maz81, bar92, fis94, pir98, ric03, geh04}). We now know that in fact GRBs encompass different kinds of physical events. It is understood that GRBs (see \cite{zha19} for a more complete review) are {\bf extreme} explosions at cosmological distances due to either the merging of two compact objects \cite{eic89}, or the death of super-massive stars \cite{woo93}. Soft gamma-ray repeaters \cite{maz82}, or tidal disruption events (such as Swift J164449.3+573451, \cite{bur11}), are also visible in gamma-rays and could be wrongly identified as GRBs.

While these powerful explosions are followed by a long lasting afterglow, visible for several years for the brightest of them (e.g. \cite{dep16}), what is powering them, the central engine, has already died and stopped after the initial part of the phenomenon \cite{pir99,mes06}. Or at least this was supposed before the launch of the Swift satellite. In fact, very early after the launch, it was discovered a plateau phase, linking the prompt and the afterglow phases \cite{nou06}. This was the first evidence that maybe the central engine was active for a longer time.

A few years later, a puzzling GRBs has been identified, GRB 111209A \cite{gen13}. This event had a duration lasting far longer than the normal GRBs, and has been dubbed an ultra-long event (hereafter ulGRB, for ultra-long Gamma-Ray Burst). Various authors have spotted several bursts with similar properties (e.g. \cite{lev14, cuc15, lie16, dew23}), inferring what has been defined as a new sub-class of event. It was not the first time a new sub-class was added to the population of GRBs. Historically, the first two were the classes of long and short events, detected by PHEBUS and confirmed by BATSE from the whole population of GRBs \cite{dez92, kou93}. \cite{muk98} added an intermediate class of events, located between short and long events, which nature is still not clear. Finally, the observations of GRB980425 and GRB 031203 allowed to define the class of low-luminosity GRBs \cite{saz04,lia07}. Hence, the creation of the new sub-class of ultra-long GRB was the most logical answer to a GRBs with very peculiar properties. While the nature of these ultra-long events is still not certain, all proposed explanations are expecting the central engine to be active for a longer time. 

In this short review, we will focus on the rare occurrences where the central engine seems to be long lived, namely on the ultra-long GRBs and the plateau phase of the afterglow. we will first address the difficulty of making accurate measurements of a duration. we will then review the observational properties of long lasting phenomena within GRBs known so far, before developing the theoretical considerations which are currently investigated.

\section{Measuring a duration with high energy experiments}

Conceptually, long lasting activities within a GRB have always been present. It is their discovery which occurred recently. One should first wonder why these phenomena have not been acknowledged with the discovery of GRBs, in the late 60's \cite{kle73}. Indeed, measuring the duration of an event was possible with the first Vela satellites \cite{kle73} and has been routinely done since then (e.g. \cite{maz81, kou93}). 

The first possible answer is that the method used to measure the duration, the $T_{90}$ measurement, was not suited for those peculiar events. Measuring $T_{90}$ implies to measure the duration where the burst emits 90\% of its observed fluence, starting at the time where 5\% of that fluence has already been emitted \cite{kos95}. Various bursts have been observed presenting an advanced emission of energy, not covering those 5\% criteria, sometime tens or hundred of seconds before the main event \cite{kos95b, bur08}. These small emissions were called precursors, and, if taken into account, would have dramatically increased the duration of the burst they were associated with \cite{li_22}. However, this was not the case, and the studies focused on the main prompt event and their afterglows: these events were not classified as peculiar at that time. Some other bursts were presenting a more convincing hint of long lasting activity in the BATSE 4B catalog, with larger duration \cite{pac99}, but none of them where classified as peculiar at that time. Hence, it is important to discuss how to understand the measurement of a duration, and how this is linked to the central engine.

\begin{figure}[H]
\centering\includegraphics[width=12 cm]{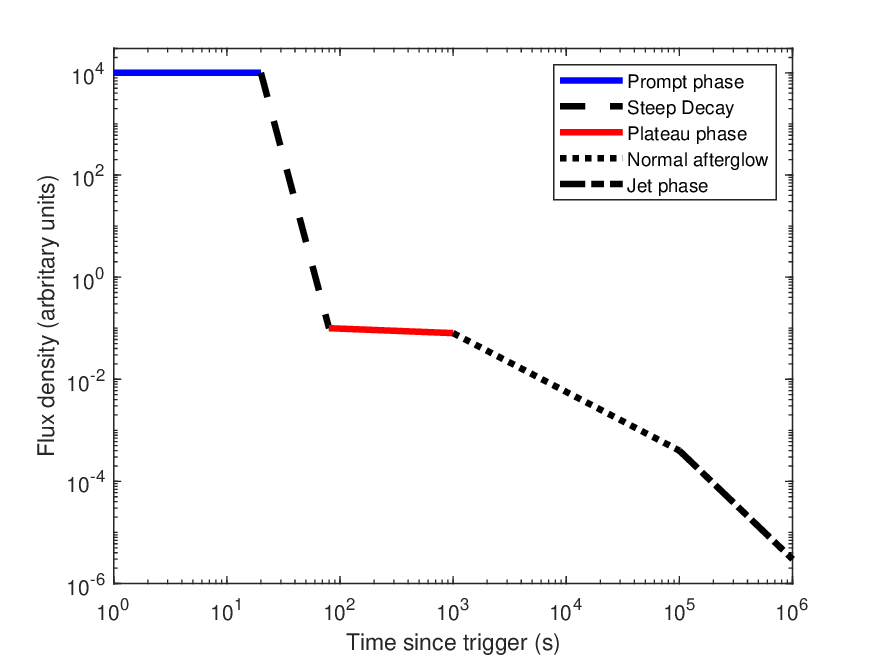}
\caption{Schematic view of a normal long gamma-ray burst lightcurve. From the left to the right the various phases are the {\it prompt phase}, the {\it steep decay phase}, the {\it plateau phase}, the {\it normal afterglow phase}, and the {\it post-jet break phase}. Within this review, we will concentrate on the two phases indicated by the solid colored lines.\label{fig1}}
\end{figure}

As a first evident point, one can measure the duration of a phenomenon only when both the start and the end of said phenomenon can be observed. Within Fig. \ref{fig1}, we have displayed a typical light curve of a normal long GRB, showing the very large duration of the whole phenomenon. Because gamma-ray detectors are located on low-orbits, the Earth is located in about one third to half of their respective field of view \cite{bro92}. Hence, the observation of any phase, and most of all the prompt phase will not be continuous in the case of long events, and there is a strong bias toward events lasting significantly less than the orbit of the satellite \cite{hak98}. In the case of events lasting more than 1000 seconds, there is a fair probability that the start or the end of the prompt phase occurs while the burst is occulted. The same applies to X-ray observations, also done by satellites. In addition, the localization precision of the previous generations of gamma-ray detectors does not allow to certify that an occulted event observed on two different orbits is indeed the same event, and not two unrelated but close by events \cite{wan95}.

Secondly, there is no clear signal of the end of the prompt phase in the gamma-ray band. The signal simply fade out into the background of the instrument, and cannot be detected anymore. In some cases, the burst can even seem to fade away into the background, and then starts again after some quiescent time (e.g. \cite{lan18,pat18}). This, however, does not carry a physical meaning (the prompt phase is over), but an observational one (the signal is too faint) \cite{sha11}. One can then wonder if these events really stopped (and potentially re-started), or were only undetectable. This is the kind of considerations which caused \cite{zha09} to built a more robust classification between short and long GRBs, based on several observational properties (association with a supernova, classification of the host, estimation of the star formation rate, offset from the host nucleus, local density of the surrounding medium, among other criteria), rather than just one parameter. This however remove the bias of measurement of the duration, but also the measurement itself: the classification of \cite{zha09} provides a physical meaning, but no more a meaningful measurement of the duration. To counter this lack, \cite{str13} and \cite{zha13} have forged independently an X-ray estimator of the duration of the activity of the central engine, based on the position of the start (for \cite{str13}) or the end (for \cite{zha13}) of the steep decay phase (dashed line in Fig \ref{fig1}), better suited for the study of ulGRBs. It is based on the works of \cite{kum00} and \cite{wil07}, who linked the steep decay phase to the high-latitude emission of the prompt phase discovered by \cite{nou06}, visible only when the on-axis emission is turned off, and the afterglow emission has not yet reach a brightness enough to mask it out. As it, the start of this phase corresponds to the end of the central engine activity, and the end of this phase corresponds when any emission related to the central engine activity has ended.

The situation is similar for features present in the afterglow phase, like the plateau phase \cite{nou06}. This phase has been missed by BeppoSAX \cite{boe97} due to its late observation of the burst with its narrow instruments \cite{cos97}. The extrapolations of the afterglow emission up to the prompt emission were compatible \cite{dep06}, albeit some hints were already present that something may be lying in between the prompt and the afterglow phases, as some events, like GRB 000214, had their afterglow observation extrapolation clearly not compatible with the prompt observations \cite{sof04}. For those reasons, most of the work done on long-lasting activities of the central engine have been done in the Swift era: only fast repointing satellites allow for a sampling of the light curve dense enough for studying these features.


\section{Observations}
\subsection{Ultra-long Gamma-Ray Bursts}

\begin{figure}[H]
\centering\includegraphics[width=12 cm]{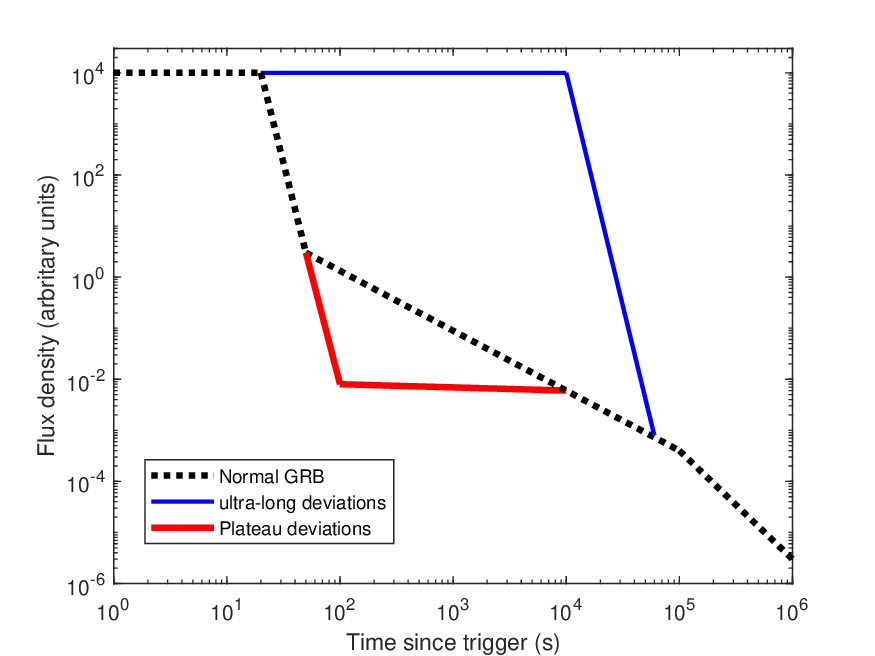}
\caption{Schematic comparison of the light curves of a normal long GRB (dashed line) with the deviations caused by an ultra-long GRB (solid top line), and a GRB with a plateau phase (solid bottom line). For simplicity, the initial prompt and the late afterglow luminosities have been fixed to a common value. This simplification is not representative of the diversity of the observed fluxes for those events, but enhance the difference between those three kind of events.\label{fig2}}
\end{figure}

The prompt phase of the {\em ultra-long} GRBs last more than $10^3$ seconds, as displayed on Fig. \ref{fig2}. As discussed in \cite{zha14} and \cite{boe15}, this time mark, used to discriminate ulGRBs and long GRBs is not well defined, and various authors are using an ad-hoc separation value, complicating the comparison of the samples. \cite{lie16} for instance use emission lasting more than 1000 seconds in the $15-350$ keV energy range. \cite{boe15} used the same duration, but in the $0.2-10$ keV band. It is thus safe to consider the value of $10^3$ seconds as an indication. Following the statistical work of \cite{gen19}, events with a prompt X-ray duration of more than 5 000 seconds in the X-ray band are for sure ulGRBs, and those with a duration of more than 1 000 seconds {\it may be} ulGRBs.

This classification has been challenged by \cite{vir13}, who claimed that ulGRBs are rather the tail of the distribution of normal long GRBs. \cite{zha13} obtained a similar conclusion using the sample of event available at that time: there was no evidence for a different origin for ulGRBs. As explained above, these authors modeled the overall light curve in X-rays (0.3 - 10 keV) and used a theoretical model linked the end of the steep decay phase to the behavior of the prompt emission at that energy to measure the end time of the prompt phase, and thus the duration. However, with a larger (and more statistically significant) sample, \cite{boe15} demonstrated that the distribution of the duration of ulGRBs was not consistent with the one of long GRBs, and since then the ulGRB class has been well established and used for further classification of new events (e.g. \cite{dew23}).

The first studies done on those events were on the properties of their afterglow phase. \cite{str13} has shown that the afterglow of ultra-long GRB 111209A is not very different from any long burst one. This result was also obtained by \cite{lev14}, using another member of the class: GRB 101225A. This is not difficult to assume when one compare all afterglow observations taken so far (the so-called Kann plot, \cite{kan24}), with their luminosity distribution spanning over several decades of energy: it is easy to fit any small discrepancy within such a broad distribution. Interestingly, for some ulGRBs it has been reported the presence of a thermal component within the spectrum \cite{pir14}, but again the same has been reported for long events \cite{ryd09,ryd10,pee12}.

At the very end of the afterglow phase, long GRBs are known to be associated with supernovae \cite{hjo03}. Ultra-long GRBs have been also associated with supernovae \cite{gre15}. Hence, on that aspect, both seems similar.

Once the afterglow has faded away, it is possible to have a view of the host galaxy. Studies such as the localization of the events with respect to the host galaxy (e.g. \cite{fon13}) have confirmed a different progenitor nature for the short and long GRBs. In the case of ulGRBs, this kind of study is difficult due to the small apparent size of the host galaxies \cite{lev14}. Their host galaxies appear different than those of normal long GRBs, albeit not unseen before \cite{lev14}. Their metalicity have been measured to be sub-solar \cite{lev14} to super-solar \cite{sch15}, indicating that the metalicity of the host may not be linked to the nature of these events.

At the opposite of the light curve, \cite{gen19} has shown that in their prompt phase, ulGRBs are similar to long GRBs in all aspects but their duration, albeit \cite{mar15} hinted for some soft events. It thus seems that the only thing defining an ultra-long GRB is its ultra-long duration, as summarized in Table \ref{tab1}.

\begin{table}[H] 
\caption{Comparison of the properties of an ultra-long GRB with a normal long GRB.\label{tab1}}
\newcolumntype{C}{>{\centering\arraybackslash}X}
\begin{tabularx}{\textwidth}{CCCCC}
\toprule
\textbf{}	        & \textbf{Typical duration (s)}	& \textbf{Typical prompt properties}\textsuperscript{1} & \textbf{Typical redshift}\textsuperscript{2} & \textbf{Presence of supernova}\\
\midrule
Normal long GRB		& 20			& $\alpha = 1.8 \pm 0.2$, $F_{-9} = 14 \pm 12$ & 2.8 & Yes\\
Ultra-long GRB    & $>5000$ &	$\alpha = 1.8 \pm 0.4$, $F_{-9} = 16 \pm 19$ & 1.6 & Yes\\
\bottomrule
\end{tabularx}
\noindent{\footnotesize{\textsuperscript{1} $\alpha$ is the spectral index, $F_{-9}$ is the mean flux in units of $10^{-9}$ erg cm$^{-2}$ s$^{-1}$. See \cite{gen19}.}}
\noindent{\footnotesize{\textsuperscript{2} As discussed in \cite{gen13}, the redshift distribution of ultra-long GRBs is biased due to the trigger methods. The value indicated for the normal long GRBs is extracted from \cite{jak06}.}}
\end{table}

\subsection{The plateau phase}

Before the launch of Swift, the observations done by BeppoSAX of the prompt and the late afterglow were nicely aligning together with a single power law decay \cite{dep06}. Things changed when \cite{nou06} studied the generic X-ray light curve of the afterglow using {\it Swift}. They find that the prompt event is rapidly followed by a steep decay, a break into a shallow decay phase, and a second break into a "normal" decay phase, with possible flares superimposed on both phases (see also Fig. \ref{fig2}, depicting such a situation, with a comparison with a normal long GRB). As it, the plateau has a very vague definition, as it is defined as a flatter portion of the light curve observed in between two steeper decays; this definition already make it difficult to provide mean properties. Its decay index is usually less than 0.7, with in exceptional cases a negative value (indicating an increase of the luminosity with time), while the afterglow decay index is usually closer to 1.2. 

Interestingly, the X-ray plateau phase is not strongly correlated with the optical emission \cite{zan13}. Optical plateaus do exist \cite{pan11}, but are not always occurring at the same time than the X-ray plateau \cite{gen12}, albeit in some cases they are possibly {\bf related to the same mechanism} \cite{dai20}. On the spectral side, its spectrum is compatible with no or few evolution between the plateau and the following normal decay emission \cite{vau06, lia07b}.

A lot of works have been done to link the properties of GRBs together and to use them as standard candles. While the first attempt by \cite{boe00} was at the end inconclusive, several well established 'relations' have survived the observation tests: the Amati relation \cite{ama02}, the Ghirlanda relation \cite{ghi04}, or the Frail relation \cite{fra01} to cite a few of them. The plateau phase has not been forgotten, and this led to the discovery of a relation between the date of the end of the plateau phase and its luminosity at that time \cite{dai13, dai16}: the Dainotti relation, which states that $L_{(t_{end pl})} \propto t_{end pl}^{-\gamma}$ (the shortest the plateau phase, the brightest the normal afterglow is when it starts). Some more analyzes seems to indicate that some extra parameters (such as the addition of the isotropic equivalent energy) may tighten these correlations (e.g. \cite{si_18}).

\section{The Central engine and the Fireball model}

Once the observational facts listed in the previous sections were established, ones started to work on explaining them. The beauty of the current canonical model for the GRBs is its versatility. Technically, what we are observing is an explosion and its aftermath. The Fireball model \cite{ree92,mes97,pan98} explains all the aspects of this explosion through the expansion of an ultra-relativistic fireball of plasma and its interactions with the surrounding medium. The genius signature of the model is that the description of the explosive, and what trigger the explosion, is hided into a black box called the central engine \cite{ree92}. The nature of long-lasting phenomena within GRBs has to be linked with one of those two components.

The central engine is in charge of producing the amount of energy needed by the fireball for its evolution (or requested to fit the observations!). With a total energy $E_{iso}$ reaching up to $\sim 10^{54}$ ergs \cite{att17}, it has been acknowledged very early in the study of the phenomenon that the central engine has to be a black hole in formation \cite{pac86}. There are currently three kinds of central engine considered for GRBs:

\begin{itemize}
\item A merging binary of neutron stars \cite{eic89} \footnote{Note a semantic point here. The progenitor of a GRB is the object which will collapse into a black hole, while the central engine is the phenomenon of the formation of the black hole. Hence, a binary of neutron stars is a progenitor, while the merging itself is the central engine}. This kind of event has been directly observed once with the association of GW170817 with GRB170817A \cite{abb17}, and is thought to produce short events \cite{zha19}.
\item The {\bf collapse} of an hypermassive magnetar \cite{uso92}. In this case, the energy is stored as the rotational energy of the magnetar, and is extracted up to the point where the magnetar cannot anymore sustain its stability by the centrifugal force, and collapse into a black hole.
\item The coalescence of a massive star, or a collapsar \cite{woo93}. This last kind of central engine is linked to the end-point of life of the massive stars, like supernovae. The presence of a large tank of matter available for the black hole to accrete can produce events lasting several seconds, explaining the nature of most of long GRBs \cite{zha19}.
\end{itemize}

One should note that it is possible that the central engine change its nature, for instance starting as the coalescence of a massive star, but then transiting by an intermediate state where a magnetar is formed first \cite{tro07}.

Energy is extracted as long as the central engine is active. Once the black hole is formed and accretion has stopped, the central engine is supposed to be turned off. There is a vivid debate to understand if it is possible to revive the central engine, for instance with a late episode of accretion, or if does not qualify as "central engine activity" \cite{kin05,mu_16,dal17,duq22}.

\begin{figure}[H]
\includegraphics[width=13.85 cm]{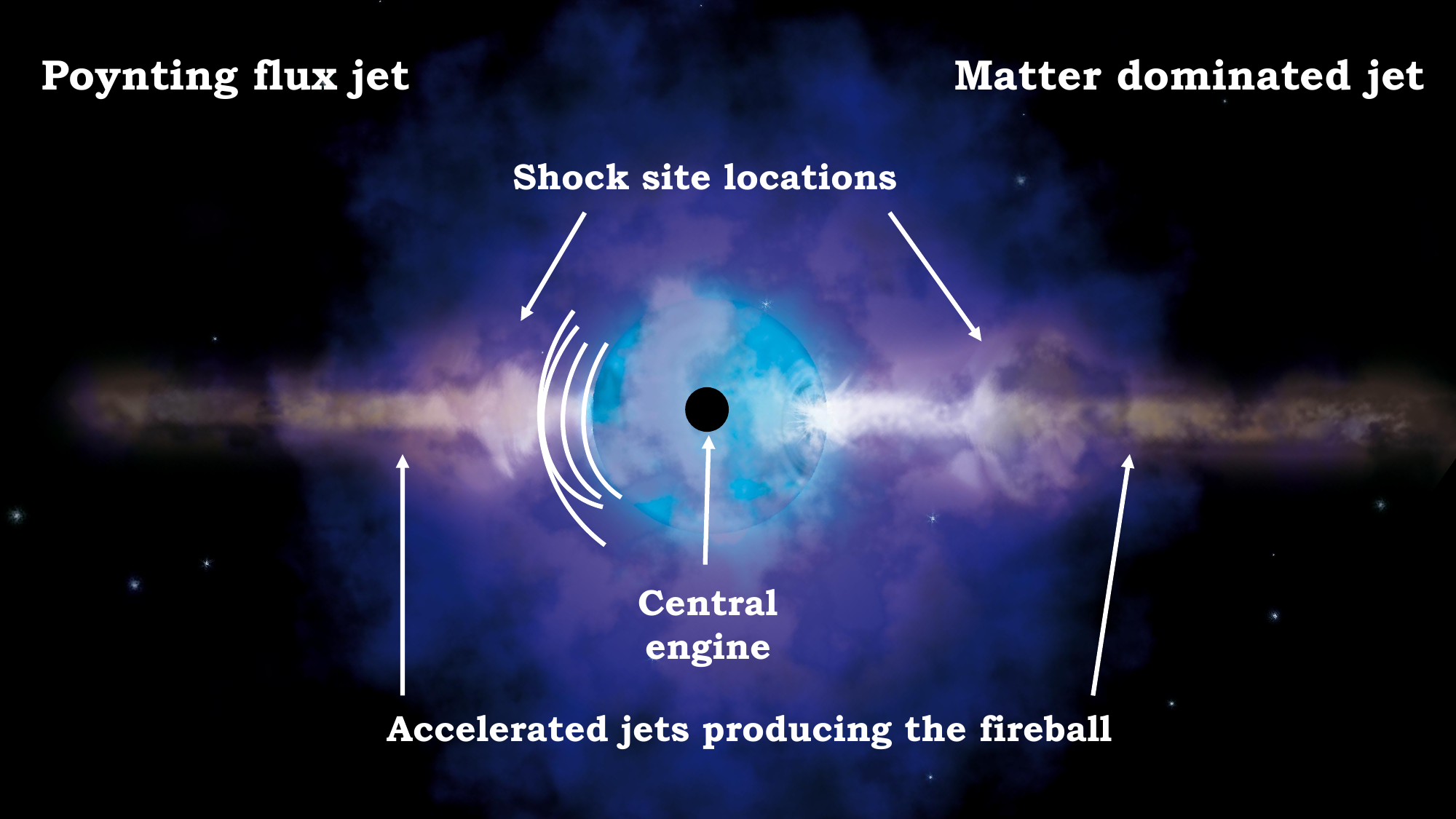}
\caption{Schematic description of the canonical model, focusing on the inner components of the model. Represented are the progenitor (as a blue supergiant star), the central engine (as the black dot at the center of the progenitor), and the jets. On the right part of the schema, the jet is a matter dominated jet where the energy extracted from the central engine is stored as kinetic energy into the jet. On the left part of the schema, the jet is a Poynting flux dominated jet, where the energy is stored inside the electromagnetic field (represented by the white field lines), and extracted when the field reconnects (this can occur at various distances, represented on the schema by the dim clear area between the progenitor and the main shock location). In both cases, some of the energy is then injected into shocks, while other can continue to travel for later shocks. The schema is simplified, and not to scale.\label{fig3}}
\end{figure}

The extraction mechanism (and the transportation of the energy to the injection point) is still debated. At the close vicinity of the central engine, the energy is extracted as heat and photons/electromagnetic field, through the electromagnetic field interaction of the magnetar, or the accretion mechanism \cite{pir99}. Then there are two extreme options: either the energy is brought directly to the injection point via Poynting flux \cite{dai98, lyu03}, or is transformed by pair-production into particles and stored as kinetic energy \cite{goo86, pac86, she90, pac90}. Obviously, intermediate scenario where these two hypotheses are mixed together do exist (e.g. \cite{mes97b, gia06, gao15}). These two different channels will in any case interact differently with the surrounding medium, injecting their energy into it. This complete mechanism is illustrated in Fig. \ref{fig3}.

All this energy is injected into the fireball, which produces the prompt and afterglow emissions characteristic of a GRB. Within the original Fireball model, it was supposed that the energy injection occurred at the start of the phenomenon and was instantaneous \cite{bla76,ree92}. With the discovery of X-ray flares, this framework has changed and now the global model can account for late energy injection \cite{pan98,dai98}. In any case, the fireball, through different kinds of shocks, both within the jet and with the surrounding medium, will dissipate this energy, until it is fully decelerated and cannot radiate anymore. 

The cooling down of the fireball and the theoretical light curves of the phenomenon can be described theoretically \cite{sar98,sar99}. Doing so, the break times and slopes of those light curves can be linked to physical properties of the fireball \cite{pan00, pan01}, while this is not the case for long lasting constant activity of the prompt or the early afterglow \cite{pan00, pan01}. Deviations from those theoretical light curves are thus very interesting, as they may indicate a signature of the central engine manifesting at late time, and thus carrying valuable information about the central engine and the progenitor: all explosions have a tendency to mask out the nature of the explosive.

There is however a last complication to this global picture. Because GRBs are a jetted phenomena, the viewing angle plays an important role. This has been theorized long ago, with the possibility to observe so-called orphan afterglows \cite{rho97, hua02, nak02}, namely GRBs where the prompt emission was missed due to their off-axis orientation. The detection and subsequent observations of GRB 170817A \cite{abb17, abb17b, gol17} have confirmed that the orientation was indeed modifying some observed properties, such as the peak time of the afterglow \cite{gra18}, which implies another theoretical stage for the models to be compared with the observations \cite{rho99, sar99, ros02, rya20}. These modifications are not trivial, as they depends on various models for the jet structure and geometry, and the possible effect of the surrounding medium, among other factors. The interested reader should refer to \cite{sal22} for a more detailed review of this peculiar topic.

\section{Explaining the long lasting activities within Gamma-Ray Bursts}

\subsection{Considerations on the nature of the progenitor and of the central engine}

There is still a vivid debate about the nature of the central engines (and thus their progenitors) to explain those long lasting features. As shown in the previous section, there are few observable properties directly linked to the central engine (such as its position within the host galaxy) to study \cite{lev15, iok16}, and by construction most of the models are predicting similar spectro-temporal observations (and the usual derived measures such as spectral and temporal indices, $E_p$, fluxes): only a detailed case by case analysis of a very complete data set can reveal the subtle differences among those model (an interesting discussion is given by \cite{iok16}). Ultra-long events can be explained with three main classes of progenitors: an ultra-massive stellar progenitor, very similar to Pop III stars \cite{suw11, nag12, mac13}; the tidal disruption of a dwarf star \cite{mac14}; and a newborn magnetar \cite{gre15}. Long lasting plateau phases are better explained by the late extraction of rotational energy stored into a magnetar \cite{tro07}, by a continuous energy injection at the start of the forward shock \cite{zha06}, or by refreshed energy injection due to the nature and geometry of the jet \citep{geng13, sun24}.

For ultra-long events, the original hypothesis of \cite{gen13} was that the progenitor should be linked to a collapsar, but with a larger star. \cite{pir14} was also arguing for a more massive progenitor. Their main arguments were that, given the similarity between long and ultra-long GRB properties, one should have the same broad class of progenitor: a massive star. End-to-end simulations, modeling the evolution and collapse of supergiant stars, done by \cite{per18} or by \cite{sun17} indicated that such a progenitor was indeed possible. In such a case, the plateau phase and the ultra-long duration of some prompt events would not be related.

Another possibility, which would relate these two kind of phenomena together is the magnetar hypothesis of \cite{gre15}. While there were claims that the energy budget of ulGRBs was too large for a magnetar hypothesis \cite{pir14}, and that the energy budget of a spinning magnetar is by definition limited\footnote{A simple calculus can show that a sub-millisecond magnetar would have its equatorial rotation speed to exceed the speed of the light, and that centripetal forces would have destroyed the object before that limit is reached.} \cite{sun24}, other studies, such as those of \cite{iok16, can16, liu18,che24} have indicated that a magnetar option was indeed possible and cannot be ruled out by the current observations. On the other hand, since the first paper of \cite{tro07}, numerous studies, such as the ones of \cite{lyo10,row13,gom14,str18b} have also pointed out a magnetar explanation of the plateau phase of gamma-ray bursts. \cite{cor09} has even studied this hypothesis with multi-messenger observations. All of these studies led to the emergence of a model for this progenitor \cite{lu_15}, and the use of the observations to study the properties of the magnetar itself \cite{row14}. In such a case, the ultra-long gamma-ray bursts and some plateau phases would share the same progenitor.

There were also studies which indicated that some plateau phases could be due to continuous energy injection within the fireball \cite{dai98,zha01}. As emphasized above, both Poynting-flux-dominated flow \cite{dai98} or pairs-dominated flow \cite{gen16} could explain this phenomenon, but would also provide very similar observations difficult to discriminate from each other. Interestingly, \cite{eva14} offered an alternative hypothesis for the nature of ultra long GRBs where the long emission was not linked to the central engine, but to the surrounding medium. By analyzing the data of ulGRB~130925A they formulate the hypothesis that ulGRBs reside in very low density environments that make the ejecta decelerate at times much longer than if they were in a denser medium. Then, ultra-long GRBs and some plateau phase events would have a similar origin, but this time based on continuous energy injection within the fireball.

\subsection{Delaying the emission linked to a weak energy extraction from the central engine}

A last interesting question to review is how late can the energy extracted from the central engine be radiated by the fireball, and is this qualify as late activity. Flares are common in the X-ray light curves of afterglows \cite{fal07,but07,li_12} and are sometime observed very late \cite{kum22}. Albeit they are considered as some energy injection from the central engine, their interpretation as witness of the central engine activity is still under debate \cite{laz07,laz11,swe13,swe14,yi_22}. Indeed, flares can be due to a renewed or continuing activity of the central engine \cite{zha06}, or to a refreshed internal shock (e.g. \cite{zhu23}). In the later case, some slow blobs of matter ejected by the central engine far earlier finally catch up with the main forward shock. As a matter of consequence, the time of the last X-ray flare could measure the velocity of the slowest shells rather than the duration of the central engine activity. Hence the conclusion that all central engines are active far longer than initially though (e.g. \cite{zha06,laz07}) may not be true.

\section{Future Perspectives and Conclusions}

Launched now 20 years ago, the Swift satellite has revolutionized the science of GRBs by showing the extended activity of the central engine. However, its design was done years before, and is not the best suited to continue the studies. {\it SVOM} \cite{gon18, wei16}, which has been launched recently, is better armed for those studies. In the case of ulGRBs, its pointing strategy will allow observing the same direction for long periods (typically several hours), and can use an image trigger extending to at least 20 minutes. This will simplify efforts to detect and recognize ulGRBs \cite{dag20}. The multiwavelength capability of {\it SVOM} will allow the prompt emission to be monitored simultaneously in the visible (GWAC), hard X-rays (ECLAIRs), and gamma-rays (GRM), while for the afterglow emission this will be in NIR and visible (GFTs and VT), and in X-rays (MXT)---providing detailed diagnostics of any continued activity of the central engine. It should also have enough sensitivity to monitor the plateau phase in great details.

If the mission get selected by ESA, THESEUS \cite{ama18} will also allow for observations within a very wide energy band (0.3 keV $-$ 20 MeV), with an unprecedented large sensitivity. It will provide very good resolution of the spectra and light curves, allowing to resolve faint features \cite{cio21,str18}. Another interesting option would be to turn to radio observations. Indeed, magnetars are pulsars, and radio observations could help constraining the nature of the progenitors \cite{law19}.

On the non-photonic messenger side of the observations, GRBs have been theorized to be high-energy particle accelerators and to produce thermal neutrinos \cite{wax95, vie95, wax97}. The acceleration site is linked to the jet geometry and its composition, and offer an interesting probe to test these parameters. While current neutrino detectors have failed to detect strong signals \cite{mur22,abb23,abb24}, upgraded detectors such as IceCube \cite{aar21} could make ground breaking detections, challenging the current jet models \cite{mur22,par24}. Similar conclusions can be reached with cosmic rays and new experiments such as the CTA \cite{act11}, namely that any new observation would strongly challenge the models (e.g. \cite{mur06, alv19, moo24}).

Lastly, several mechanisms discussed in this review are also gravitational wave emitters. Therefore, the next generation of instruments (the Einstein Telescope, NEMO, and LISA, \cite{pun10,ack20,ama17}), which will have an horizon encompassing the mean distance of ulGRBs \cite{pun10}, may be able to provide some more clues about the exact phenomenon we are observing, using multi-messenger studies.

In any case, the studies about long lasting events within gamma-ray bursts are clearly not complete yet, and the current and future missions are suited to continue them. Thus, it would not be surprising if new results would appear within the next decade, and help completing their current understanding.

\vspace{6pt} 


\funding{This research received no external funding.}

\dataavailability{No new data were created or analyzed in this study. Data sharing is not applicable to this article.} 

\acknowledgments{Parts of this review have been supported by the Australian Research Council Centre of Excellence for Gravitational Wave Discovery (OzGrav), through project number CE170100004.}

\conflictsofinterest{The authors declare no conflicts of interest.}

\begin{adjustwidth}{-\extralength}{0cm}

\reftitle{References}

\bibliography{review}

\PublishersNote{}
\end{adjustwidth}
\end{document}